\documentclass[aps,prl,twocolumn,superscriptaddress,showpacs]{revtex4}

\usepackage{longtable}
\usepackage{graphicx}
\usepackage{dcolumn}
\usepackage{bm}
\usepackage{amsmath}
\usepackage[psamsfonts]{amssymb}

\newcommand{\STO}{SrTiO$_3$}
\newcommand{\LAO}{LaAlO$_3$}
\newcommand{\AFM}{antiferromagnetism}
\newcommand{\MR}{magnetoresistance}

\newcommand{\etal}{\emph{et al.}}
\newcommand{\TDEG}{two dimensional electron gas}

\begin{document}

\title{Anisotropic magneto-transport effects at \STO$\backslash$\LAO~interfaces}


\author{M. Ben Shalom}
\affiliation{Raymond and Beverly Sackler School of Physics and Astronomy, Tel-Aviv University, Tel Aviv, 69978, Israel}
\author{C. W. Tai}
\affiliation{Department of Physical Electronics, School of
Electrical Engineering, The Iby and Aladar Fleischman Faculty of
Engineering, Tel-Aviv University, Tel Aviv, 69978, Israel}
\author{Y. Lereah}
\affiliation{Department of Physical Electronics, School of
Electrical Engineering, The Iby and Aladar Fleischman Faculty of
Engineering, Tel-Aviv University, Tel Aviv, 69978, Israel}
\author{M. Sachs}
\affiliation{Raymond and Beverly Sackler School of Physics and Astronomy, Tel-Aviv University, Tel Aviv, 69978, Israel}
\author{E. Levy}
\affiliation{Raymond and Beverly Sackler School of Physics and Astronomy, Tel-Aviv University, Tel Aviv, 69978, Israel}
\author{D. Rakhmilevitch}
\affiliation{Raymond and Beverly Sackler School of Physics and Astronomy, Tel-Aviv University, Tel Aviv, 69978, Israel}
\author{A. Palevski}
\affiliation{Raymond and Beverly Sackler School of Physics and Astronomy, Tel-Aviv University, Tel Aviv, 69978, Israel}

\author{Y. Dagan}
\email[]{yodagan@post.tau.ac.il} \affiliation{Raymond and Beverly Sackler School of Physics
and Astronomy, Tel-Aviv University, Tel Aviv, 69978, Israel}


\date{\today}

\begin{abstract}
The resistivity as a function of temperature, magnetic field and its orientation for atomically flat
\STO$\backslash$\LAO~ interfaces with carrier densities of $\sim$ 3$\times$10$^{13}$ cm$^{-2}$ is reported.
At low magnetic fields superconductivity is observed below 130mK.
The temperature dependence of the high field \MR~ and
its strong anisotropy suggest possible magnetic ordering below 35K.
The origin of this ordering and its possible relation to superconductivity are discussed.
\end{abstract}
\pacs{75.70.Cn, 73.40.-c }
\maketitle
Interface between strongly correlated electron materials can be
very different from their constituents. It has been
shown that if \LAO~ is epitaxially grown on TiO$_2$-terminated
\STO~ a \TDEG~ is formed at the interface between these insulators
\cite{OhtomoHwang}. This interface was latter shown to be
superconducting \cite{ReyrenSC} and magnetic \cite{BrinkmanSC}.
Recently Caviglia \etal~ have shown that the superconducting
transition temperature can be controlled by solely varying the
number of charge carriers at the interface using a gate
voltage.\cite{CavigiliaGating} These unexpected results and the
potential for development of high performance oxide based
electronics motivated an effort to understand the properties of
this interface \cite{WillmottPRL, HerranzPRL, Zhicheng_Zhong} and
to improve it.
\par
The origin of the large carrier concentration at the interface
remains under debate.  When depositing monolayers of \LAO~ on
\STO~ conductivity appears only for a TiO$_2$ terminated surface
\cite{OhtomoHwang} at a threshold of 4 unit cells.\cite{Thiel}
These observations suggest that the electrostatic structure of the
interface: nonpolar \STO~ planes covered with alternatingly
charged planes on the \LAO~ side should lead to an interfacial
reconstruction. This reconstruction can be dominantly electronic
in nature,\cite{OkamotoMillis, popovictheoryfortwodeg} or partly
due to cationic mixing.\cite{WillmottPRL} A lattice distortion
driven by the polar nature of the interface has also been
proposed.\cite{pentchevaPicketPRL} Other papers suggested that
oxygen vacancies play a major role in creating high carrier
densities. \cite{kalabukhovOVac, HerranzPRL,siemons:196802} It
seems that the latter effect is insignificant for samples
deposited at pressure range of $10^{-5}$ to $10^{-3}$
Torr.\cite{Nakagawa_no_O_defects, ReyrenSC, Huijben_multilayer}
\par
Magnetic effects have been theoretically predicted for
\STO$\backslash$\LAO~ interfaces.\cite{Zhicheng_Zhong,
pentcheva:035112} Recent observations of magnetic hysteresis below
0.3K along with magneto-resistance oscillations with periodicity
proportional to $\sqrt{B}$ have been explained in terms of
commensurability of states formed at the terrace edges of the
\STO~ substrate.\cite{MOscillations}
\par
While superconductivity in this interface has been shown to be two
dimensional in nature \cite{ReyrenSC} the way such interface can
exhibit magnetic properties is still a puzzle. In this paper we
show that for carrier concentrations of $3\times10^{13} cm^{-2}$
the \TDEG~ is superconducting at 130mK, yet, novel
magneto-transport effects are observed below 35K. Our data support
possible evidence for a magnetic order formed below this
temperature. A magnetic impurities scenario is ruled out.
\begin{figure}
\includegraphics[height=0.9\hsize]{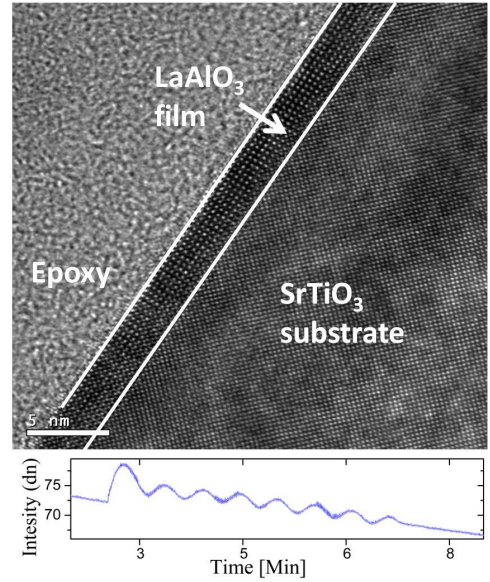}
\caption {(a)High resolution transmission electron microscopy
image of \LAO$\backslash$\STO interface. The lines outline the
\LAO~ film boundaries. The number of layers is as expected from
the number of RHEED oscillations. (b) RHEED oscillationsindicating
a film deposition of thickness of 8 unit cells. \label{RHEEDTEM}}
\end{figure}
\par
Eight unit cells of \LAO~ were deposited from a single crystal
target onto a TiO$_2$-terminated \STO~ prepared in a similar way
as described by Koster \etal~\cite{kosteridealstosurface} by
pulsed laser deposition. We use pulse rate of 1Hz and energy
density of 1.5 J$\times$cm$^{-2}$ at oxygen pressure ranging
between 1$\times10^{-3}$ - 5$\times10^{-5}$ Torr and temperature
of $800^{\circ}C$. The deposition was monitored by reflection high
energy electron diffraction (RHEED). The maxima of the RHEED
intensity oscillations indicate a complete layer formation and
used as a measurement for the sample thickness (see
Fig.~\ref{RHEEDTEM}b.). One of the samples was imaged by a high
resolution transmission electron microscope revealing a high
quality interface and confirming the thickness measurement by the
RHEED (see Fig.~\ref{RHEEDTEM}a.). The \TDEG~ underneath the \LAO~
layers was electrically connected using a wire bonder. One of the
samples was patterned using reactive ion etch (RIE) into Hall bars
with bridges dimensions of 50*750 microns squared. The bridges
were align perpendicular or parallel to the terrace edges. Other
samples were connected in a Van-Der Pauw (VDP) geometry for
resistivity and Hall measurements, or in a strip geometry (with
dimensions of about $2mm\times0.1mm$) when the current direction
had to be well defined.
\par
In this paper we present four typical samples deposited at oxygen
pressures of $5\times10^{-5}$ (Sample 1), $1\times10^{-4}$
(Samples 2 and 4) and $9\times10^{-4}$ (Sample 3), with carrier
concentrations of $3\times10^{13}cm^{-2}$,
$5\times10^{13}cm^{-2}$, $2\times10^{13}cm^{-2}$ and
$3.5\times10^{13}cm^{-2}$ for samples 1-4 respectively as inferred
from Hall measurements at 2K. The charge carrier density has a
very weak temperature dependence up to 100K. This is in contrast
with the strong temperature dependence reported in
Ref.\cite{BrinkmanSC}.
\par
\begin{figure}
\includegraphics[width=1\hsize]{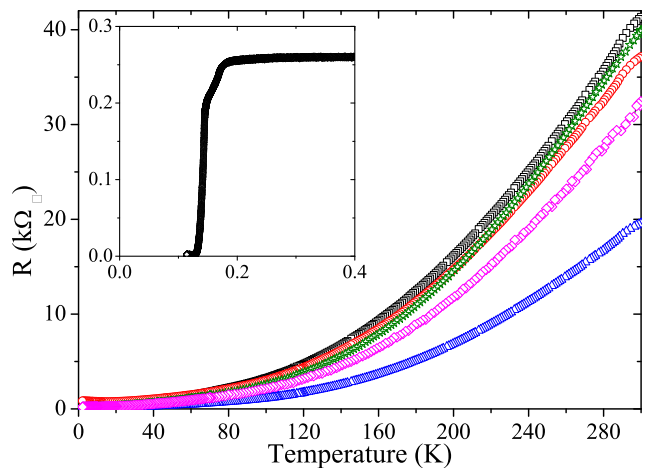}
\caption {(Color online) The sheet resistance as a function of
temperature for three typical samples of \STO~$\backslash$  8 unit
cells \LAO~ interface. Sample1 (black squares) sample 2 (red
circles) sample3 (blue triangles) and the two bridges of sample 4
(green stars and magenta diamonds). Insert: sheet resistance
versus temperature for sample 1.\label{RTandPT}}
\end{figure}

The sheet resistance as a function of temperature for these
samples is shown in Fig.~\ref{RTandPT}. All samples under study
including all bridges in the patterned sample exhibit similar
transport properties. The fact that small bridges and VDP
measurements resulted in similar features is indicative of the
samples' homogeneity. We also note that the variation of oxygen
pressure during deposition resulted in a rather small change in
carrier concentration and resistivity. Sample 1 was also measured
in a dilution refrigerator and was shown to be superconducting
with the transition temperature T$_c$=130mK (see insert of
Fig.\ref{RTandPT}).
\par
\par
\begin{figure}
\includegraphics[width=1\hsize]{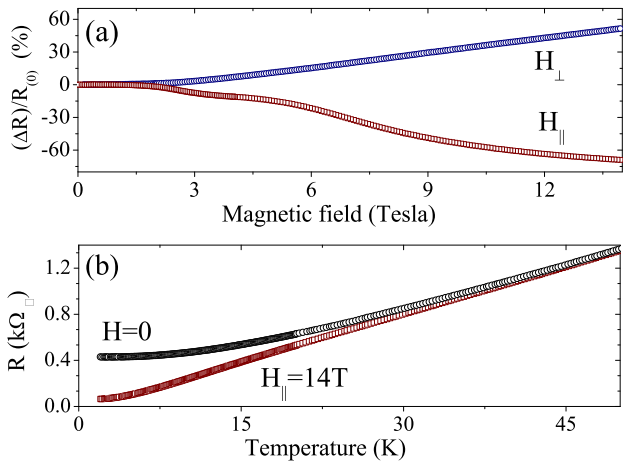}
\caption {(Color online), Sample1 (a) Blue circles: the \MR~ as a function
of magnetic field applied perpendicular to the interface. Red
squares are the \MR~ data for field applied along the interface
parallel to the current. (b) The sheet resistance as a function of
temperature at zero field (black circles) and at 14 Tesla applied
parallel to the current (red squares)\label{MR}}
\end{figure}

The magnetoresistance (MR) is defined as $\frac{\Delta
R}{R_0}=\frac{R(H)-R(H=0)}{R(H=0)}$, where R(H) is the resistance
at a magnetic field H. It is presented for T=2K (Sample1) in
Fig.~\ref{MR}(a). When the field is applied perpendicular to the
film a positive MR is observed (blue circles). The data is an
average between positive and negative fields in order to eliminate
spurious Hall contribution due to contact misalignment. By
contrast a large negative MR is seen for fields parallel to the
film and to the current (red squares). We note that both the
positive and negative MR are very large, 50\% and 70\%
respectively for a magnetic field of 14 Tesla. We also note that
for perpendicular fields no hysteresis is observed down to 130mK
where superconductivity shows up.
\par
In Fig.~\ref{MR}(b)we show the temperature dependence of the
(parallel) negative MR (Sample1). The black circles are the zero
field measurement and the red circles are data taken at 14 Tesla
applied parallel to the current. We emphasize that the negative MR
disappears above 35K. The large negative MR and its strong
anisotropy suggest strong magnetic scattering in the plane. To
further investigate this assumption we rotated the field around a
horizontal axis changing its angle with the normal to the
interface while keeping the field's amplitude constant (14 Tesla).
\par
\begin{figure}
\includegraphics[width=1\hsize]{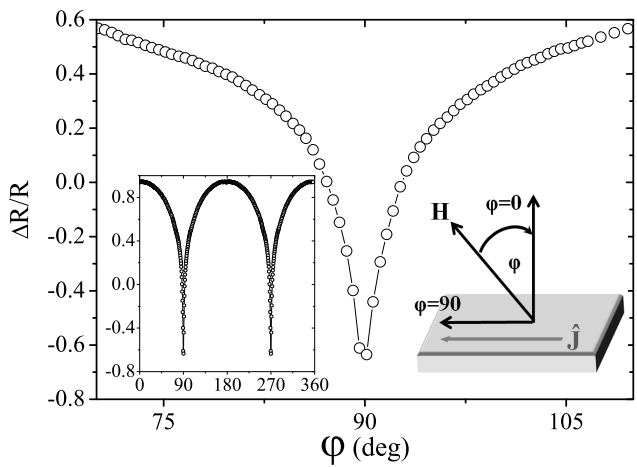}
\caption {Sample2, the \MR~ as a function of the angle
$\varphi$ between the perpendicular to the interface and the
magnetic field ($\varphi$ is depicted at the right insert). Left
Insert: full angle scan.\label{peprotation}}
\end{figure}

In Fig.~\ref{peprotation} the MR for sample2 at 14 Tesla is
plotted as a function of the angle between the magnetic field and
the normal to the film (see insert for illustration). $90^\circ$
corresponds to a magnetic field applied parallel to the current.
The dip is extremely sharp (see insert for the entire angle scan)
and the MR changes sign at $87^\circ$ (or $93^\circ$).
\par
\begin{figure}
\includegraphics[width=1\hsize]{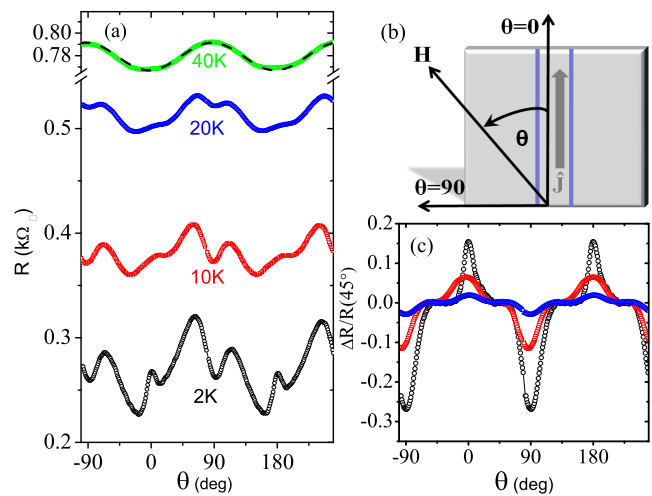}
\caption {(Color online) Sample3 (a) The \MR~ at 14 Tesla as a function of
angle between the field and the current at various temperatures
(data points). The dashed line is a $sin^2(\theta)$ fit (see text
for details) (b) Illustration of the measurement geometry. (c) The
same data as in (a) after subtracting the fit and normalizing with
the measured resistance at $\theta=45^{\circ}$. This procedure
uncovers the (in-plane) anisotropic \MR~.\label{Rotation}}
\end{figure}

We shall now check for anisotropy in the plane of the interface.
In Fig.\ref{Rotation} the resistance as a function of angle
between the magnetic field and the current is shown for various
temperatures (Sample3). $\theta=90^{\circ}$ corresponds to a magnetic
field applied parallel to the film and perpendicular to the
current (see illustration Fig.\ref{Rotation}(b)). At 40K the
resistance is maximum for a field perpendicular to the current.
The dashed black line Fig.\ref{Rotation}(a) is a fit using
$R(\theta,T)=r(T)sin^2(\theta)+R_0(T)$, with $\theta$ being the angle
between the field and the current, $r(T=40K)=25 \Omega$ and
$R_0(T=40K)=767 \Omega$. This simple dependence persists up to above
100K. As elaborated in the discussion below we attribute this
behavior to geometric effects related to the two dimensional
nature of the electron gas at the interface.
\par
Below 40K another effect appears. Focusing for example on the 20K
data, a dip appears at $\pm90^{\circ}$ while a peak is revealed at
0, $180^{\circ}$. At 10K and below the latter effect becomes even
larger than the $sin^2(\theta)$ and a significant maxima (minima)
appears at $0^{\circ}$ and $180^{\circ}$ ($90^{\circ}$ and
$-90^{\circ}$). We note that since the interface is probably not
perfectly parallel to the field a small Hall contribution results
in a small deviation between zero and $180^{\circ}$ Moreover such
a small deviation can result in a perpendicular component,
although this component is minute its influence can be
non-negligible and should add-up to the $sin^2(\theta)$ effect. To
eliminate the Hall contribution we symmetrized the data for
positive and negative fields. To remove the contributions with the
$sin^2(\theta)$ dependencies we subtracted the fit
$R(\theta,T)=r(T)sin^2(\theta)+R_0(T)$ from the 2K, 10K and 20K
data $r(T)$ and $R_0(T)$ were determined for each temperature.
This procedure uncovers the (in-plane) anisotropic \MR~. The
resulting data normalized with the measured resistance at
$\theta=45^{\circ}$ are shown in Fig.\ref{Rotation}(c). We note a
sharp peak when the field is parallel to the current and a sharp
dip appears when the field is applied perpendicular to it. A
similar effect is seen for a different strip rotated by 90 degrees
(not shown). As elaborated below we interpret this effect as being
the anisotropic MR. A small deviation between the angles at which
the maximum appears for various temperatures could be due to
different angular sweep direction and the rotator backlash.
\par
We shall now discuss the MR data from Figs.\ref{MR},
\ref{peprotation}, \ref{Rotation}. We first note that the
amplitudes of both negative and positive MR in Fig.\ref{MR}a. are
very large. The positive MR for a field perpendicular to the
interface could be due to orbital effects that have a significant
contribution since $\omega_c\tau$ is close to unity, where
$\omega_c$ is the cyclotron frequency and $\tau$ the scattering
time. For a field parallel to the current such orbital effects are
not existent. Yet, the MR is even larger, ~70\%. The relevant
mechanisms that can produce negative MR are: two dimensional (2D)
weak localization, magnetic impurities and the magnetic nature of
the material itself. The first effect is ruled out since it is
usually small (of the order of a few percents) and appears for
field applied perpendicular to the film. The second effect is
usually isotropic, in strong contrast with our results.  We are
therefore led to conclude that the large negative MR we observe is
due to a magnetic order formed at the interface.
\par
We emphasize that the negative MR seen in Fig.\ref{MR}a. for
$\varphi=90^{\circ}$ is very different from the negative isotropic
MR reported in Ref.\cite{BrinkmanSC}. The MR versus $\varphi$
dependance shown in Fig.\ref{peprotation} is extremely sharp
around $\varphi=90^{\circ}$. This is a key observation in our
paper. The fact that the MR changes sign for a variation of
$3^{\circ}$ implies that a small perpendicular field component is
sufficient to mute the mechanism responsible for the parallel
negative MR. This is due to the fact that when
$\varphi=93^{\circ}$ the parallel field component is almost
unchanged (13.98 Tesla) while the perpendicular component is only
0.73 Tesla. Such a component is too small to induce any orbital
effect as can be seen in Fig.\ref{MR}a. One may claim that the
positive orbital MR for 0.73 Tesla is in fact larger, yet
overwhelmed by a large isotropic negative MR. However, when
measuring the negative MR with a parallel field of 0.73 Tesla we
find it to be very small (see . Hence this scenario is ruled out.
We therefore conclude that there is a strong strange anisotropy of
the MR. The only element in our system with such strong
directionality is the interface itself. We therefore conclude that
the strong $\varphi$ dependence gives possible evidence for the
existence of magnetic order confined to a few layers near the
interface. This magnetic order in the interface vanishes above 35K
according to the data in Fig.\ref{MR}b. for the carrier density
and \LAO~ thickness under study.
\par
Further evidence for the quasi-2D nature of the
conducting interface can be found from the in-plane angular
dependence of the MR as presented in Fig.\ref{Rotation} at 40K.
For this geometry (field and current in plane) the Lorentz force
is perpendicular to the interface. Assuming a quasi 2D confinement
one expects an enhancement of scattering as the field is applied
perpendicular to the current assuming that the band structure is
not very simple. This \emph{positive} orbital contribution to the
MR should be quadratic in the field component that is
perpendicular to the current. We observed a $sin^2(\theta)$
behavior as expected (see dashed line Fig.\ref{Rotation}).
\par
We can roughly estimate the width of the confinement zone using a
naive calculation with the mean free path at low temperatures
$\ell=\frac{h}{e^2k_FR_\square}\approx 25nm$ at 2K, the Fermi wave
number $k_F=\sqrt{2\pi n_s}$, $e$ the electron charge, $R_\square$
the sheet resistance, and $n_s$ the carrier density. The ratio
between this MR and the one observed when the field is applied
perpendicular to the interface should be proportional to
$(d/\ell)^2$ where $d$ is the size of the confinement zone.
Substituting the values for $R_\square$ and the amplitude of the
two orbital effects at 40K we obtain $d\approx1-2nm$. This gives
the right order of magnitude for the width of the confinement
zone. We note that this effect and all other effects reported here
are similar for current running parallel or perpendicular to the
substrate terraces, which rules out the terraces as their origin
in contrast with ref.\cite{MOscillations}.
\par
In summary, we have deposited sharp \STO~$\backslash$\LAO~
interfaces. We studied the temperature field and orientation
dependence of the MR. We identify four contributions to the MR:
(a) an orbital one, measured when the field is perpendicular to
the interface, (b) the $sin^2(\theta)$ MR persisting up to rather
high temperatures. This MR appears when the field is applied
parallel to the interface and perpendicular to the current. We
relate it to the finite size of the confinement zone. This MR is
also positive, but its amplitude comparing to the previous effect
is smaller by a factor proportional to $(d/\ell)^2$. (c) The more
interesting MR appears below 35K. This, negative, low temperature
MR appears when the field is applied exactly parallel to the
interface and to the current. It cannot be due to orbital effects
and its large (negative) magnitude suggests that it has a magnetic
origin. (d) the last effect is seen when rotating the field in
plane. Below 35K anisotropic MR appears. It has a maximum for
$\theta=0$ (parallel to the current). Its amplitude increases as
the temperature decreases. We interpret this MR as being the
anisotropic MR expected for magnetic
materials.\cite{AnisotropicMRreview} Scattering resulting from
spin-orbit interactions becomes stronger when the electron travels
parallel to the magnetization as seen in Fig.\ref{Rotation}(c).
The latter two effects: the strong (parallel) negative MR and the
anisotropic, in-plane MR show up \emph{together} below 35K. Below
this temperature a magnetic phase emerges. This phase is extremely
sensitive to an out of plane magnetic field. This sensitivity is
unclear to us, yet, it rules out magnetic impurities as the origin
of the effects and suggests that the magnetic order is confined to
the vicinity of the interface. We take note of the following
observations: both the parallel negative MR and the anisotropic,
in-plane MR  exhibit no saturation up to 14 tesla, and we were not
able to observe magnetic hysteresis down to T$_c$=130mK. In view
of these observations and due to the occurrence of
superconductivity at low temperatures it is difficult to believe
that the interface is ferromagnetic. We speculate that
antiferromagnetic order is formed at the interface below a
N\`{e}el temperature of 35K. this temperature may vary with number
of charge carriers and film thickness.  Providing that this \AFM~
is not induced by the applied field this system could be another
example of coexistence of antiferromagnetism and superconductivity
as in heavy fermion materials \cite{Mathur} and in some of the
cuprates.\cite{daganResistivityPRL}

\begin{acknowledgments}
We thank A. Aharony, G. Deutscher, A. Gerber,  S. Hacohen-Gourgy,
O. Entin-Wohlman and Y. Yacoby for useful discussions. This
research was partially supported by the Israel Science Foundation
(ISF) by the FIRST program of the ISF by the Binational Science
Foundation and by the Wolfson family charitable trust. CWT is
grateful for the financial support by Grant No. 29637 from
European Commission.
\end{acknowledgments}

\bibliographystyle{apsrev}
\bibliography{STOLAO}
\end{document}